\documentclass[12pt]{article}
\usepackage{graphicx}
\usepackage{color}
\usepackage{cite}
\def\downstrut{\vrule height 1ex depth 1.0ex width 0pt}
\def\upstrut{\vrule height 2.5ex depth 0.0ex width 0pt}

\def \beq{\begin{equation}}
\def \eeq{\end{equation}}
\def\eqref#1{(\ref{#1})}
\def\bea{\begin{eqnarray}}
\def\eea{\end{eqnarray}}

\def\URLtilde{\lower0.2em\hbox{$\tilde{\phantom{a}}$}}
\def\mycomm#1{\hfill\break\strut\kern-3em{\color{red}\tt ====> #1
\color{black}}\hfill\break}

\newcount\timecount
\newcount\hours \newcount\minutes  \newcount\temp \newcount\pmhours
\hours = \time
\divide\hours by 60
\temp = \hours
\multiply\temp by 60
\minutes = \time
\advance\minutes by -\temp
\def\hour{\the\hours}
\def\minute{\ifnum\minutes<10 0\the\minutes
\else\the\minutes\fi}
\def\clock{
\ifnum\hours=0 12:\minute\ AM
\else\ifnum\hours<12 \hour:\minute\ AM
\else\ifnum\hours=12 12:\minute\ PM
\else\ifnum\hours>12
\pmhours=\hours
\advance\pmhours by -12
\the\pmhours:\minute\ PM
\fi
\fi
\fi
\fi
}

\def\monthname{\relax\ifcase\month 0/\or January\or February\or
March\or April\or May\or June\or July\or August\or September\or
October\or November\or December\else\number\month/\fi}

\def\bold#1{\setbox0=\hbox{$#1$}     \kern-.025em\copy0\kern-\wd0
\kern.05em\copy0\kern-\wd0
\kern-.025em\raise.0433em\box0 }

\begin{document}
\setcounter{footnote}{1}
\rightline{EFI 18-19}
\rightline{TAUP 3034/18}
\rightline{arXiv:1906.07799}
\vskip1.0cm

\centerline{\large \bf Status of isospin splittings in mesons and baryons}
\bigskip

\centerline{Marek Karliner$^a$\footnote{{\tt marek@proton.tau.ac.il}}
 and Jonathan L. Rosner$^b$\footnote{{\tt rosner@hep.uchicago.edu}}}
\medskip

\centerline{$^a$ {\it School of Physics and Astronomy}}
\centerline{\it Raymond and Beverly Sackler Faculty of Exact Sciences}
\centerline{\it Tel Aviv University, Tel Aviv 69978, Israel}
\medskip

\centerline{$^b$ {\it Enrico Fermi Institute and Department of Physics}}
\centerline{\it University of Chicago, 5620 S. Ellis Avenue, Chicago, IL
60637, USA}
\bigskip
\strut

\begin{quote}
\begin{center}
ABSTRACT
\end{center}
Current measurements of isospin splittings in mesons and baryons are
sufficiently precise that they allow estimates of the mass difference
between constituent up and down quarks.  Some previous results are updated
in the light of these new measurements, and the importance of better
measurements of some observables such as $M(K^{*\pm})$, $M(B^{*0})-M(B^0)$,
and isospin splittings in bottom baryons is noted.
\end{quote}
\smallskip

\leftline{PACS codes: 14.20.Lq, 14.20.Mr, 12.40.Yx}
\bigskip


\section{INTRODUCTION \label{sec:intro}}

Isospin-violating mass differences among hadrons are treated in the quark model
as a combination of effects.  The $u$ and $d$ quarks have an intrinsic mass
difference, expressed as a direct contribution to hadron masses and via
differing kinetic energies in bound states.  Coulomb interactions between
quarks depend on the product of their charges times the expectation value of
the inverse of their separation.  Strong hyperfine interactions between quarks
depend on the inverse product of their masses, and electromagnetic hyperfine
interactions depend both on that inverse product and on the product of quark
charges. One can then write meson and baryon isospin-violating mass differences
in terms of a few parameters, yielding sum rules for masses in the limit of
small values of these parameters. These were exploited, for example, for mesons
with heavy quarks in Ref.\ \cite{Rosner:1992qw} and for baryons in Ref.\
\cite{Rosner:1998zc}.  Isospin splittings in baryons with two heavy quarks
were examined in Refs.\ \cite{Brodsky:2011zs} and \cite{Karliner:2017gml}.

The experimental status of isospin splittings continues to improve.  There has
been a relatively new measurement of $M(D^{*+})-M(D^+)$ \cite{TheBaBar:2017yff}.
Information on masses of individual charge states of charmed and bottom hadrons
continues to grow, with exceptional progress in the past year for $\Xi^c$,
$\Sigma_b$, and $\Xi_b$ states \cite{Tanabashi:2018oca} (compare PDGLive with
the 2018 print version).  An update of Ref.\ \cite{Rosner:1992qw} was performed
about ten years ago \cite{Goity:2007fu}.  Even the light-quark sector has seen
improvements since the analysis of Ref.\ \cite{Rosner:1998zc}, driven by the
improved precision in the $\Xi^0$ mass measured by the NA48 Collaboration at
CERN \cite{Fanti:1999gy}.  An analysis of the present status of isospin
splittings in hadrons thus seems appropriate.

We set forth our assumptions, including the interpretation of quarks as
constituents with masses of several hundred MeV, in Section \ref{sec:a}.
In Sec. \ref{sec:q} we update analyses of light-quark mesons and baryons.
We treat charmed hadrons in Sec.\ \ref{sec:c}, beauty hadrons in Sec.\
\ref{sec:b}, and the relation between the two heavy sectors in Sec.\
\ref{sec:bc}.  We compare our results with those of several other approaches
in Sec.\ \ref{sec:comp}, and conclude in Sec.\ \ref{sec:concl}.

\section{ASSUMPTIONS \label{sec:a}}

In a constituent-quark framework, hadron masses are governed by the sum of
their quark masses, the hyperfine interactions among those quarks, and --- for
hadrons with more than one heavy quark ($c$ or $b$) --- an additional binding
term between heavy quarks.  This approach \cite{DeRujula:1975qlm,%
Lipkin:1978eh} successfully describes the masses of light-quark hadrons
\cite{Gasiorowicz:1981jz}, those with a single charm or bottom quark
\cite{Karliner:2014gca}, and the mass of the recently observed baryon with
two charmed quarks \cite{Aaij:2017ueg}.

When the masses of mesons and baryons are fitted with constituent-quark
masses and hyperfine interactions, the quark masses in baryons are about 55
MeV heavier than those in mesons \cite{Lipkin:1978eh}.  This scheme was used
in Ref.\ \cite{Karliner:2014gca} to predict $M(\Xi_{cc}) = (3627 \pm 12)$ MeV,
in satisfactory agreement with the observed value \cite{Aaij:2017ueg}
$M(\Xi_{cc}) = (3621.40 \pm 0.78)$ MeV.  An alternative scheme explains the
mass difference by adding a ``string-junction'' term of 165 MeV, allowing
one to fit mesons and baryons with a universal set of quark masses
\cite{Karliner:2016zzc}.  However, this scheme predicts $M(\Xi_{cc})$ about
40 MeV higher, so for definiteness we shall stay with the picture of separate
quark masses for mesons and baryons.

Quark masses in \cite{Karliner:2014gca}, from Ref.\ \cite{Gasiorowicz:1981jz},
did not include a small binding term for a pair of $s$ quarks, which we now
take into account.  The results are shown in Table \ref{tab:fit}.  Here the
strong hyperfine term is parametrized as
\beq \label{eqn:hfs}
\Delta E_{ij,{\rm HFs}}=b\langle \sigma_i \cdot \sigma_j \rangle /(m_i m_j)~.
\eeq
Superscripts $m$ and $b$ will refer to values in mesons and baryons,
respectively.  The quark masses differ only slightly from those in Ref.\
\cite{Karliner:2014gca}.

\begin{table}
\caption{Results (in MeV) of a fit to mesons and baryons with additive quark
masses (different for mesons and baryons), hyperfine terms, and a binding
term $B(ss)$.  The label $\bar m$ denotes an average between $m_u$ and $m_d$.
The masses of $\phi$, $\Xi$, and $\Omega$ are corrected by terms $-2B(ss),
~-B(ss),$ and $-3B(ss)$, respectively.  We find $\bar m^m = 307.5$ MeV, $m^m_s
= 487.6$ MeV, $\bar m^b = 362.1$ MeV, $m^b_s = 543.9$ MeV, $b^m/(\bar m^m)^2 =
79.4$ MeV, $b^b/(\bar m^b)^2 = 50.0$ MeV, $B(ss) = 9.2$ MeV.  The
root-mean-square error of the fit is $\sqrt{\sum(\Delta M^2)/13} = 3.85$ MeV.
\label{tab:fit}}
\begin{center}
\begin{tabular}{c c c c c c} \\ \hline \hline
Meson & $\pi$ & $\rho$ & $K$ & $K^*$ & $\phi$  \\ \hline
Pred. & 138.5 & 773.9 & 494.6 & 895.3 & 1019.9 \\
Expt. & 138.0 & 775.2 & 495.6 & 894.1 & 1019.5 \\
$\Delta M^2$ & 0.2 &  1.8  &  1.0  &  1.4  & 0.2 \\ \hline \hline
\end{tabular}
\vskip 0.1in

\begin{tabular}{c c c c c c c c c} \hline \hline
Baryon & $N$ & $\Delta$ & $\Lambda$ &
 $\Sigma$ & $\Sigma^*$ & $\Xi$ & $\Xi^*$ & $\Omega$ \\ \hline
Pred.& 936.5 & 1236.4 & 1118.2 & 1185.0 &
 1385.0 & 1329.7 & 1529.4 & 1670.4 \\
Expt. & 938.9 & 1232. & 1115.7 & 1193.2 & 1384.2 & 1321.0 & 1532.5 & 1672.5 \\
$\Delta M^2$ & 6.0 & 18.9 & 6.2 & 67.3 & 0.2 & 75.3 & 9.9 & 4.5 \\
\hline \hline
\end{tabular}
\end{center}
\end{table}

The model we employ takes into account the intrinsic difference $\Delta =
(1-\frac{K}{m})(m_u - m_d)$ between $u$ and $d$ quarks, where $K$ is a one-body
kinetic energy term \cite{Isgur:1979ed}; Coulomb interactions
\beq
\Delta E_{ij~{\rm em}} = \alpha Q_i Q_j \langle 1/r_{ij} \rangle
\eeq
between quarks; strong hyperfine (HF) interactions $\Delta E_{ij~{\rm HFs}}$
as mentioned above; and electromagnetic HF interactions
\beq
\Delta E_{ij~{\rm HFe}} = - \frac{2 \pi \alpha Q_i Q_j|\Psi_{ij}(0)|^2 \langle
\sigma_i \cdot \sigma_j \rangle}{3m_i m_j}~.
\eeq
Symbols are defined in Ref.\ \cite{Rosner:1998zc}.  We may thus write
the total isospin splitting as
\beq
\sum_{i<j} \Delta E_{ij}=\langle \Delta \rangle + a \sum_{i<j} \langle Q_i Q_j
\rangle + b \sum_{i<j} \langle \sigma_i \cdot \sigma_j/(m_i m_j) \rangle
+ c \sum_{i<j} \langle Q_i Q_j \sigma_i \cdot \sigma_j/(m_i m_j) \rangle~.
\eeq
Separate parameters, labeled by superscripts $m$, will be used for mesons.
Henceforth parameters without superscripts will refer to quantities for baryons.

\section{LIGHT-QUARK HADRONS \label{sec:q}}

\subsection{Mesons}

The isospin splittings of light-quark mesons, based on masses quoted in
Ref.\ \cite{Tanabashi:2018oca}, are summarized in Table \ref{tab:lqm}.
Labels denote the change in isospin associated with each mass splitting.
The conflict between the quoted $K^*$ mass splitting (which we use) and the
individual $K^*$ masses needs to be resolved before we can take our analysis
as definitive.

\begin{table}
\caption{Isospin splittings of light-quark mesons \cite{Tanabashi:2018oca}.
\label{tab:lqm}}
\begin{center}
\begin{tabular}{c c c} \hline \hline
$\Delta M$ & Label & Value (MeV) \\ \hline
$\pi^\pm - \pi^0$ & $\pi_2$ & 4.5936$\pm$0.0005 \\
$\rho^\pm - \rho^0$ & $\rho_2$ & 0.15$\pm$0.42 \\
$K^+ - K^0$ & $K_1$ & --3.934$\pm$ 0.020 \\
$K^{*+} - K^{*0}$ & $K^*_1$ &--6.7$\pm$1.2$^a$ \\ \hline \hline
\end{tabular}
\end{center}
\leftline{$^a$As quoted in pdgLive.  Individual masses quoted are
891.76$\pm$0.25 MeV (charged,}
\leftline{hadroproduced); 895.5$\pm$0.8 MeV (charged, $\tau$ decay); and
895.55$\pm$0.20 MeV (neutral).}
\end{table}

Each splitting can be written as the sum of terms depending on the parameters
$\Delta^m$, $a^m$, $b^m$, and $c^m$:

\bea
\pi_2  & = & \frac{1}{2}a^m + \frac{3}{2}\frac{b^m}{(\bar m^m)^2} \left(
\frac{\Delta^m}{\bar m^m} \right)^2 - \frac{3}{2} \frac{c^m}{(\bar m^m)^2}~, \\
\rho_2 & = & \frac{1}{2}a^m - \frac{1}{2}\frac{b^m}{(\bar m^m)^2} \left(
\frac{\Delta^m}{\bar m^m} \right)^2 + \frac{1}{2} \frac{c^m}{(\bar m^m)^2}~, \\
K_1 & = & \Delta^m + \frac{1}{3}a^m + \frac{3b^m}{(\bar m^m)^2}
\frac{\Delta^m}{m^m_s} - \frac{c^m}{(\bar m^m)^2}\frac{\bar m^m}{m^m_s}~,\\
K^*_1 & = & \Delta^m + \frac{1}{3}a^m - \frac{b^m}{(\bar m^m)^2}
\frac{\Delta^m}{m^m_s} + \frac{c^m}{3(\bar m^m)^2}\frac{\bar m^m}{m^m_s}~.
\eea
Here we have substituted $\Delta^m$ for $m^m_u-m^m_d$ in the terms for
hyperfine splittings.  We then see that four observables are expressed in
terms of the three unknowns $\Delta^m$, $a^m$, and $c^m$.  (We use the value of
$b^m/(\bar m^m)^2 = 79.4$ MeV from the fit in Table \ref{tab:fit}.)  A fit to
the observables yields the values $\pi_2 = 4.593$ MeV, $\rho_2 = -0.065$ MeV,
$K_1 = -3.935$ MeV, and $K^*_1 = -3.200$ MeV, with $\Delta^m = -4.117$ MeV,
$a^m = 2.119$ MeV, and $c^m/(\bar m^m)^2 = -2.315$ MeV.  The $\chi^2$ for this
fit is 8.771, nearly all (8.508) contributed by the $K^*_1$.  In view of the
non-unanimity of the Particle Data Group's values for $M(K^{*\pm})$
\cite{Tanabashi:2018oca}, we urge further study of this state.

\subsection{Baryons}

We express the observed mass splittings among the octet baryons and the
$\Sigma^*$ and $\Xi^*$ resonances \cite{Tanabashi:2018oca}, labeled with
subscripts denoting their $\Delta I$ values, summarized in Table \ref{tab:lqb},
as functions of $\Delta$ ($u-d$ mass difference with effect on
kinetic energies), $a$ (Coulomb interaction), $b$ (strong HF interaction),
and $c$ (electromagnetic HF interaction).  We have neglected effects of
two-body kinetic energy operators and additional small corrections
\cite{Isgur:1979ed}.  This decomposition is summarized below, where we
have linearized expressions from Ref.\ \cite{Karliner:2017gml} in $\Delta$.
Here $\bar m$ is the average of $m_u$ and $m_d$.

\begin{table}
\caption{Experimental mass splittings between light-quark baryons
\cite{Tanabashi:2018oca}.
\label{tab:lqb}}
\begin{center}
\begin{tabular}{c c c} \hline \hline
Splitting & Symbol & Value (MeV) \\ \hline
$M(p) - M(n)$ & $N_1$ & --1.2933 \kern2.7em\\
$M(\Sigma^+) - M(\Sigma^-)$ & $\Sigma_1$ & $-8.08\kern0.45em 
\pm 0.08$\kern0.5em \\
$M(\Sigma^+) - 2M(\Sigma^0) + M(\Sigma^-)$ & $\Sigma_2$ &
$1.535 \pm 0.090$ \\
$M(\Sigma^{*+}) - M(\Sigma^{*-})$ & $\Sigma^*_1$ & $-4.40\kern0.45em 
\pm 0.61$\kern0.5em \\
$M(\Sigma^{*+}) - 2 M(\Sigma^{*0}) + M(\Sigma^{*+})$ & $\Sigma^*_2$ &
  $\kern0.45em2.6 \pm 2.1$\kern0.5em \\
$M(\Xi^0) - M(\Xi^-)$ & $\Xi_1$ & $-6.85 \kern0.45em\pm 0.21$\kern0.5em \\
$M(\Xi^{*0}) - M(\Xi^{*-})$ & $\Xi^*_1$ & $-3.20\kern0.45em\pm0.68$\kern0.5em\\
\hline \hline
\end{tabular}
\end{center}
\end{table}

\bea
N_1 &=& \Delta+\frac{a}{3}-\frac{2 b\Delta}{\bar m^3}+\frac{c}{3\bar m^2}~, \\
\Sigma_1 & = & 2 \Delta -\frac{a}{3}
  + \frac{2 b \Delta}{\bar m^3} \left[ -1 + 2 \frac{\bar m}{m_s} \right]
  + \frac{c}{3\bar m^2} \left[ 1 + 4 \frac{\bar m}{m_s} \right]
  = N_1 + \Xi_1~, \\
\Sigma^*_1 & = & 2 \Delta -\frac{a}{3}
  -\frac{2 b \Delta}{\bar m^3} \left[ 1 + \frac{\bar m}{m_s} \right]
  + \frac{c}{3\bar m^2} \left[ 1 - 2 \frac{\bar m}{m_s} \right]~, \\
\Sigma_2 & = & a + \frac{c}{\bar m^2} = \Sigma_2^*~, \\
\Xi_1 & = & \Delta - \frac{2a}{3} + \frac{4b\Delta}{\bar m^2 m_s}
  + \frac{4c}{3\bar m m_s}~,\\
\Xi^*_1 & = & \Delta - \frac{2a}{3} - \frac{2b\Delta}{\bar m^2 m_s}
  - \frac{2c}{3\bar m m_s}~,
\eea

The predicted isospin splittings are very close to the observed ones, since
the Coleman-Glashow relation \cite{CG} $\Sigma_1 = N_1 + \Xi_1$ is very close
to being obeyed by those quantities with the smallest experimental errors.  The
derived parameters are $\Delta = -2.491$ MeV, $a = 3.052$ MeV, and $c/\bar m^2
= -1.523$ MeV.  The predicted isospin splittings are compared with the observed
ones in Table \ref{tab:lq}.  The $\chi^2$ for the fit is 0.642, driven mainly
by the $\Sigma^*$ splittings.

\begin{table}
\caption{Predicted and observed isospin splittings in light-quark baryons.
\label{tab:lq}}
\begin{center}
\begin{tabular}{c r r r r r r r r} \hline \hline
 & $N_1$ & $\Sigma_1$ & $\Sigma^*_1$ & $\Sigma_2$ & $\Sigma^*_2$ & $\Xi_1$
 & $\Xi^*_1$ \\ \hline
Pred.\ & -1.293 & -8.087 & -4.685 & 1.529 & 1.529 & -6.794 & -3.392 \\ 
Expt.\ & -1.293 & -8.080 & -4.400 & 1.535 & 2.600 & -6.850 & -3.200 \\
Error  &$\sim 0$&  0.080 &  0.610 & 0.090 & 2.100 &  0.210 &  0.680 \\
$\chi^2$ & 0.000&  0.007 &  0.218 & 0.004 & 0.260 &  0.072 &  0.079 \\
\hline \hline
\end{tabular}
\end{center}
\end{table}

In comparison with Ref.\ \cite{Rosner:1998zc}, the following relation is
satisfied to greater accuracy:
\beq
\Sigma_1 - \Xi_1 (= -1.23 \pm 0.22 {~\rm MeV}) = \Sigma^*_1 - \Xi^*_1 (= -1.20
\pm 0.91 {~\rm MeV})~.
\eeq
On the other hand, the relation
\beq
\Sigma_2 (= 1.535 \pm 0.090 {~\rm MeV})=\Sigma^*_2 (= 2.6 \pm 2.1{~\rm MeV})
\eeq
is still plagued with a large experimental error on the right-hand side.

\subsection{Meson-baryon comparison}

The parameters $\Delta$, $a$, and $c/\bar m^2$ derived from fits to isospin
splittings in mesons and baryons are compared in Table \ref{tab:mbcomp}.
The signs are consistent, but central values are rather different.  Slightly
different parameters are obtained if one adopts a model in which quark
masses are universal for mesons and baryons \cite{Karliner:2017gml}.  The
difference between parameters obtained from mesons and baryons is not
surprising in view of the large spread of values for $m_u-m_d$ obtained in
various models (see Sec. IV in \cite{Rosner:1998zc}).

\begin{table}
\caption{Parameters describing isospin splittings in light-quark mesons and
baryons. \label{tab:mbcomp}}
\begin{center}
\begin{tabular}{c c c c} \\ \hline \hline
      & $\Delta$ & $a$ & $c/\bar m^2$ \\ \hline
Meson & --4.12 & 2.20 & --2.32 \\
Baryon &--2.49 & 3.05 & --1.52 \\ \hline \hline
\end{tabular}
\end{center}
\end{table}

\section{CHARMED HADRONS \label{sec:c}}

\subsection{Mesons}

\begin{table}
\caption{Masses and isospin splittings of charmed mesons, in MeV.
\label{tab:charm}}
\begin{center}
\begin{tabular}{c c} \hline \hline
$D^+$ & $1869.65 \pm 0.05$ \\
$D^0$ & $1864.83 \pm 0.05$ \\
$D_1 \equiv M(D^+)-M(D^0)$ & $4.822 \pm 0.015^a$ \\
$D^{*+}$ & $2010.26 \pm 0.05$ \\
$D^{*0}$ & $2006.85 \pm 0.05$ \\
$D^*_1 \equiv M(D^{*+})-M(D^{*0})$ & $3.41 \pm 0.07$ \\ \hline \hline
\end{tabular}
\end{center}
\leftline{$^a$Value given separately in $D^0$ section of
\cite{Tanabashi:2018oca}.}
\end{table}

The states at our disposal are summarized in Table \ref{tab:charm}.
There is not enough information to derive a set of parameters describing
these mass differences.  However, the total spin-dependent terms contribute
in a manner proportional to $\langle \sigma_i \cdot \sigma_j \rangle$, so
one may write
\bea \label{eqn:cm} D_1 & = & -\Delta^m_c + \frac{2a^m_c}{3} - 3h^m_c~, \\
D^*_1 & = & -\Delta^m_c + \frac{2a^m_c}{3} + h^m_c~,
\eea
where the superscript denotes charmed mesons.  Eliminating the hyperfine
contribution $h^m_c$, one finds
\beq \label{eqn:Dacm}
-\Delta^m_c + \frac{2a^m_c}{3} = 3.76~{\rm MeV}~.
\eeq
This is to be compared with the corresponding value for light-quark mesons,
\beq
-\Delta^m + \frac{2a^m}{3} = (4.12 + 1.41)~{\rm MeV} = 5.53~{\rm MeV}~.
\eeq
As for the hyperfine term $h^m_c = 0.35$ MeV, it contains both strong $b^m_c$
and electromagnetic $c^m_c$ contributions, which cannot be separated from
one another without further assumptions.

\subsection{Baryons}

In analogy for the light-quark baryons, we write expressions for
isospin splittings of charmed baryons:

\bea
\Sigma_{c1}  & = & 2 \Delta_c + \frac{5}{3} a_c
  + \frac{2b_c}{\bar m^2} \frac{\Delta_c}{\bar m} \left[ -1
  + 2 \frac{\bar m}{m_c} \right] + \frac{c_c}{3\bar m^2}
  \left[ 1 - \frac{8 \bar m}{m_c} \right] \\
\Sigma^*_{c1} & = & 2 \Delta_c + \frac{5}{3} b_c
  - \frac{2b_c}{\bar m^2}\frac{\Delta_c}{\bar m} \left[ 1
  + \frac{\bar m}{m_c} \right] + \frac{c_c}{3\bar m^2} 
  \left[ 1 + \frac{4 \bar m}{m_c} \right] \\
\Sigma_{c2}  & = & a_c + \frac{c_c}{\bar m^2} = \Sigma^*_{c2} \\
  \Xi_{c1}   & = & \Delta_c + \frac{1}{3} a_c
  + \frac{3b_c}{\bar m^2}\frac{\Delta_c}{m_s}
  + \frac{c_c}{\bar m m_s} \\
\Xi'_{c1}  & = & \Delta_c + \frac{1}{3} a_c
  + \frac{b_c \Delta_c}{\bar m^2} \left( \frac{2}{m_c} 
  - \frac{1}{m_s} \right) - \frac{c_c}{3 \bar m} \left(
  \frac{1}{m_s} + \frac{4}{m_c} \right) \\
\Xi^*_{c1} & = & \Delta_c + \frac{1}{3} a_c
  - \frac{b_c \Delta_c}{\bar m^2} \left( \frac{1}{m_s}
  + \frac{1}{m_c} \right) + \frac{c_c}{3 \bar m} \left(
  \frac{2}{m_c} - \frac{1}{m_s} \right)
\eea

We update a couple of relations, noted in Ref.\ \cite{Franklin:1975yu}, which
follow from our assumptions.  In 1998 the relation
\beq
\Sigma_{c2} \equiv M(\Sigma_c^{++}) - 2 M(\Sigma_c^+) + M(\Sigma_c^0)
 = \Sigma_2~,
\eeq
appeared to be violated \cite{Rosner:1998zc}, with the left-hand side
giving $-2.0 \pm 1.3$ MeV while the right-hand side gave $1.71 \pm 0.18$
MeV.  The present status of charmed baryon masses and isospin splittings
is summarized in Table \ref{tab:charb}.  The sum rule is now satisfied,
with the left-hand side giving $1.92 \pm 0.82$ MeV while the right-hand
side gives $1.535 \pm 0.090$ MeV.
\begin{table}
\caption{Masses and isospin splittings of charmed baryons, in MeV.
\label{tab:charb}}
\begin{center}
\begin{tabular}{c c} \hline \hline
$M(\Sigma_c^{++})$ & $2453.97 \pm 0.14$ \\
$M(\Sigma_c^+)$    & $2452.9 \pm 0.4$ \\
$M(\Sigma_c^0)$    & $2453.75 \pm 0.14$ \\
$\Sigma_{c1} \equiv M(\Sigma_c^{++})-M(\Sigma_c^0)$ & $0.220 \pm 0.013^a$ \\
$\Sigma_{c2} \equiv M(\Sigma_c^{++})-2M(\Sigma_c^+)+M(\Sigma_c^0)$ &
  $1.92 \pm 0.82$ \\
$M(\Sigma_c^{*++})$ & $2518.41^{+0.21}_{-0.19}$ \\
$M(\Sigma_c^{*+})$  & $2517.5 \pm 2.3$  \\
$M(\Sigma_c^{*0})$  & $2518.48 \pm 0.20$  \\
$\Sigma^*_{c1} \equiv M(\Sigma_c^{*++})-M(\Sigma_c^{*0})$ & $0.01 \pm 0.15^b$ \\
$\Sigma^*_{c2} \equiv M(\Sigma_c^{++})-2M(\Sigma_c^{*+})+M(\Sigma_c^{*0})$ &
$1.89\pm 4.61$  \\
$M(\Xi^+_c)$ & $2467.93 \pm 0.18$ \\
$M(\Xi^0_c)$ & $2470.91 \pm 0.25$ \\
$\Xi_{c1} \equiv M(\Xi^+_c) - M(\Xi^0_c)$ & $-2.98 \pm 0.22^c$ \\
$M(\Xi'^+_c)$ & $2578.4 \pm 0.5$ \\
$M(\Xi'^0_c)$ & $2579.2 \pm 0.5$ \\
$\Xi'_{c1} \equiv M(\Xi'^+_c) - M(\Xi'^0_c)$ & $-0.8 \pm 0.6^d$ \\
$M(\Xi^{*+}_c)$ & $2645.57 \pm 0.26$ \\
$M(\Xi^{*0}_c)$ & $2646.38 \pm 0.21$ \\
$\Xi^*_{c1} \equiv M(\Xi^{*+}) - M(\Xi^{*0})$ & ${-}0.80\pm0.26^e$ \\
\hline \hline
\end{tabular}
\end{center}
\leftline{$^a$Listed in \cite{Tanabashi:2018oca}, $\Sigma_c$ section.
$^b$Listed in \cite{Tanabashi:2018oca}, $\Sigma_c^*$ section.
$^c$Listed in \cite{Tanabashi:2018oca}, $\Xi_c$ section.}
\leftline{$^d$Listed in \cite{Tanabashi:2018oca}, $\Xi'^+_c(2578)$ section.
$^e$Listed in \cite{Tanabashi:2018oca}, $\Xi_c(2645)$ section.}
\end{table}

Another sum rule \cite{Franklin:1975yu},
\beq \label{eqn:sigxi}
\Sigma_{c1} - 2 \Xi'_{c1} = \Sigma^*_1 - 2 \Xi^*_1~,
\eeq
is beginning to be tested, with the left-hand side yielding $1.8 \pm 1.2$
MeV while the right-hand side is $2.0 \pm 1.5$ MeV.  The large errors are
associated both with $\Xi'_{c1}$ and $\Xi^*_1$.  A further relation is
\beq
\Sigma^*_{c1} - 2 \Xi^*_{c1} = \Sigma^*_1 - 2 \Xi^*_1~,
\eeq
where the left-hand side is $1.61 \pm 0.54$ MeV.  The sum rule is satisfied,
with the main uncertainty coming from the right-hand side.

The information about charmed baryons is complete enough that one can perform
a fit to their isospin splittings, determining parameters $\Delta_c$, $a_c$,
and $c_c/\bar m^2$ which may be compared with their light-quark counterparts.
Fixed parameters in this fit (see the caption to Table I, with $m_c$ taken
from \cite{Karliner:2014gca}) are
\beq
\bar m = 362.1~{\rm MeV},~m_s = 543.9~{\rm MeV},~b_c/(\bar m^2) =
b/(\bar m)^2 =  50.0~{\rm MeV}, m_c = 1710.5~{\rm MeV}~.
\eeq
The results of this fit are summarized in Table \ref{tab:cq}.  The derived
parameters are $\Delta_c = -2.494$ MeV, $a_c = 2.769$ MeV, and $c_c/\bar m^2
= -0.853$ MeV. The first two are rather close to those obtained for light-quark
baryons. while the last is of the same sign but only about half as large as
$c/\bar m^2$.  The $\chi^2$ for the fit is $3.284$, driven mainly by
$\Xi^*_{c1}$.

\begin{table}
\caption{Predicted and observed isospin splittings in charmed baryons.
Errors are experimental values, used in calculating $\chi^2$ contributions.
\label{tab:cq}}
\begin{center}
\begin{tabular}{c r r r r r r r r} \hline \hline
 & $\Sigma_{c1}$ & $\Sigma_{c2}$ & $\Sigma^*_{c1}$ & $\Sigma^*_{c2}$
 & $\Xi_{c1}$ & $\Xi'_{c1}$ & $\Xi^*_{c1}$ \\ \hline
Fit    & 0.221 & 1.916 &-0.064 & 1.916 & -2.827 & -1.058 & -1.200 \\
Expt.\ & 0.220 & 1.920 & 0.010 & 1.890 & -2.980 & -0.800 & -0.800 \\
Error  & 0.013 & 0.820 & 0.150 & 4.610 &  0.220 &  0.600 &  0.260 \\
$\chi^2$&0.003 & 0.000 & 0.243 & 0.000 &  0.484 &  0.185 &  2.369 \\
\hline \hline
\end{tabular}
\end{center}
\end{table}

\section{BEAUTY HADRONS \label{sec:b}}

\subsection{Mesons}

The information on beauty mesons relevant for analysis of isospin splittings
is summarized in Table \ref{tab:bmes}.  An analysis parallel to that for
charmed mesons is not possible in the absence of a value of $M(B^{*0})$.
Thus in analogy to Eq.\ (\ref{eqn:cm}) all we can write is
\bea \label{eqn:bm} B_1 & = & \Delta^m_b + \frac{a^m_b}{3}- 3h^m_b~, \\
                  B^*_1 & = & \Delta^m_b + \frac{a^m_b}{3}+  h^m_b~,
\eea
Eliminating the spin-dependent term $h^m_b$, one finds
\beq \label{eqn:Dabm}
\Delta^m_b + \frac{a^m_b}{3} = \frac{1}{4}(B_1 + 3 B^*_1)~.
\eeq

\begin{table}
\caption{Masses and isospin splittings of beauty mesons, in MeV.
\label{tab:bmes}}
\begin{center}
\begin{tabular}{c c} \hline \hline
$B^+$ & $5279.33 \pm 0.13$ \\
$B^0$ & $5279.64 \pm 0.14$ \\
$B_1 \equiv M(B^+)-M(B^0)$ & $-0.31 \pm 0.07$ \\
$M(B^{*+}) - M(B^+)$ & $45.37 \pm 0.21$ \\
$M(B^{*+})$ & $5324.70 \pm 0.27$ \\
$M(B^{*0})$ & -- \\
$B^*_1 \equiv M(B^{*+})-M(B^{*0})$ & -- \\ \hline \hline
\end{tabular}
\end{center}
\end{table}
Now, $h^m_b$ contains quark charges different from those in $h^m_c$,
but is smaller in magnitude by about a factor of $m^m_b/m^m_c \simeq 3$.
Thus we probably make an error of only about 0.1 MeV in neglecting it.
In that case we would predict $B^*_1 \simeq B_1 \simeq -0.31\pm 0.07$ MeV.
This is consistent with the Particle Data Group's charge-averaged value
$M(B^*) - M(B) = 45.22 \pm 0.21$ MeV, to be compared with $M(B^{*+}) - M(B^+)
= 45.37 \pm 0.21$ MeV \cite{Tanabashi:2018oca}, implying that $B^*$ and
$B$ isospin splittings are not too different from one another.  Definitive
conclusions await the measurement of $M(B^{*0})$.  For light-quark mesons, 
the combination $\Delta^m + \frac{a^m}{3}$ is equal to $(-4.12 + 0.71)$ MeV =
--3.41 MeV.

\subsection{Baryons}

The relevant masses of beauty baryons are summarized in Table \ref{tab:bbar}.
Here we have only information on $\Delta I = 1$ splittings, as the neutral
$\Sigma_b$ and $\Sigma_b^*$ masses are still unmeasured.  The decomposition
of isospin splittings in terms of $\Delta_b$, $a_b$, $b_b$, and $c_b$ is:
\bea \label{eqn:bbar}
\Sigma_{b1} & = & 2 \Delta_b-\frac{1}{3} a_b
  -\frac{2 b_b\Delta_b}{\bar m^3} \left[ 1 - \frac{2 \bar m}{m_b} \right]
  + \frac{c_b}{3 \bar m^2}\left[1 + \frac{4\bar m}{m_b} \right]~,\\
\Sigma^*_{b1} & = & 2 \Delta_b - \frac{1}{3} a_b
- \frac{2b_b \Delta_b}{\bar m^3} \left[ 1 + \frac{\bar m}{m_b} \right]
  + \frac{c_b} {3\bar m^2} \left[1 - 2\frac{\bar m}{m_b} \right]~,\\
\Sigma_{b2} & = & a_b + \frac{c_b}{\bar m^2} = \Sigma_{b2}^*~, \\
\Xi_{b1} & = & \Delta_b - \frac{2}{3} a_b
  + \frac{3 b_b\Delta_b}{\bar m^2 m_s}
  +\frac{c_b} {\bar m m_s}~,\\ 
\Xi'_{b1} & = & \Delta_b - \frac{2}{3} a_b
  +\frac{\Delta_b b_b}{\bar m^2} \left[\frac{2}{m_b}-\frac{1}{m_s} \right]
  + \frac{c_b}{3\bar m} \left[ \frac{2}{m_b} - \frac{1}{m_s} \right]~,\\
\Xi^*_{b1} & = & \Delta_b - \frac{2}{3} a_b
  - \frac{\Delta_b b_b}{\bar m^2} \left[\frac{1}{m_s} + \frac{1}{m_b} \right]
  -\frac{c_b}{3\bar m} \left[ \frac{1}{m_s} + \frac{1}{m_b} \right] ~.
\eea

\begin{table}
\caption{Masses and isospin splittings of beauty baryons, in MeV.
\label{tab:bbar}}
\begin{center}
\begin{tabular}{c c} \hline \hline
$M(\Sigma_b^+)$ & $5810.56 \pm 0.25$ \\
$M(\Sigma_b^-)$ & $5815.64 \pm 0.27$ \\
$\Sigma_{b1} \equiv M(\Sigma_b^+) - M(\Sigma_b^-)$ & $-5.06 \pm 0.18^a$ \\
$M(\Sigma_b^{*+})$ & $5830.32 \pm 0.27$ \\
$M(\Sigma_b^{*-})$ & $5834.74 \pm 0.30$ \\
$\Sigma_{b1}^* \equiv M(\Sigma^{*+}_b) - M(\Sigma^{*-}_b)$ & $-4.37\pm0.33^a$\\
$M(\Xi_b^0)$ & $5791.8 \pm 0.5^b$ \\
$M(\Xi_b^-)$ & $5797.0 \pm 0.9^c$ \\
$\Xi_{b1} \equiv M(\Xi_b^0) - M(\Xi_b^-)$ & $-5.9 \pm 0.6$ \\
$M(\Xi_b^{*0})$ & $5952.3 \pm 0.9$ \\
$M(\Xi_b^{*-})$ & $5955.33 \pm 0.13$ \\
$\Xi_{b1}^* \equiv M(\Xi_b^{*0}) - M(\Xi_b^{*-})$ & $-3.03 \pm 0.91$ \\
\hline \hline
\end{tabular}
\end{center}
\leftline{$^a$From PDGLive \cite{Tanabashi:2018oca}. $^b$LHCb value
\cite{Aaij:2014esa} $^c$LHCb value 
\cite{Aaij:2014lxa}.}
\end{table}

\begin{table}
\caption{Predicted and observed isospin splittings in beauty baryons.
Errors are experimental values, used in calculating $\chi^2$ contributions.
\label{tab:bq}}
\begin{center}
\begin{tabular}{c c c c c c c c c} \hline \hline
 & $\Sigma_{b1}$ & $\Sigma_{b2}$ & $\Sigma^*_{b1}$ & $\Sigma^*_{b2}$
 & $\Xi_{b1}$ & $\Xi'_{b1}$ & $\Xi^*_{b1}$ \\ \hline
Fit    & -5.015 & 0.410 & -4.522 & 0.410 & -5.979 & -3.095 & -2.848 \\
Expt.\ & -5.060 &  --   & -4.370 &  --   & -5.900 &   --   &  3.030 \\
Error  &  0.180 &  --   &  0.330 &  --   &  0.600 &   --   &  0.910 \\
$\chi^2$& 0.063 &  --   &  0.211 &  --   &  0.017 &   --   &  0.040 \\
\hline \hline
\end{tabular}
\end{center}
\end{table}

One may perform a fit to these quantities, varying $\Delta_b$, $a_b$, and
$c_b/\bar m^2$.  Fixed parameters in this fit (see the caption to Table I,
with $m_b$ taken from \cite{Karliner:2014gca}) are
\beq
\bar m = 362.1~{\rm MeV},~m_s = 543.9~{\rm MeV},~b_b/(\bar m^2) =
b/(\bar m)^2 =  50.0~{\rm MeV}, m_b = 5043.5~{\rm MeV}~.
\eeq
The results are shown in Table \ref{tab:bq}.  The associated $\chi^2$ is
0.331, so a consistent set of parameters is obtained.  However, they
differ from those fitting the light-quark or charmed baryons:
\beq \label{eqn:sbfit}
\Delta_b = -1.561~,~~a_b = 3.197~,~~c_b/\bar m^2 = -2.788~
\eeq
Two relations analogous to those for charmed baryons are predicted:
\beq
\Sigma_{b1} - 2 \Xi'_{b1} = \Sigma^*_{b1} - 2\Xi^*_{b1} = \Sigma^*_1
- 2 \Xi^*_1~,
\eeq
with the second holding only for equal light-quark baryon and beauty baryon
parameters.  The right-hand side of this relation is
\beq
{\rm R.H.S.} = a - \frac{2 \Delta b}{\bar m^2} \left[ \frac{1}{\bar m}
 - \frac{1}{m_s} \right] + \frac{c}{3 \bar m^2} \left[1 + \frac{2\bar m}{m_s}
\right]~,
\eeq
whether for light-quark, charmed, or beauty baryons.  In Sec.\ \ref{sec:c}
we found $\Sigma^*_1 - 2 \Xi^*_1 = 2.0 \pm 1.5$ MeV.  However,
the large splitting between neutral and charged $\Xi_b$ states leads the
middle term of this sum rule to the value
\beq
\Sigma^*_{b1} - 2\Xi^*_{b1} = [-4.37 \pm 0.33 + 2(5.9 \pm 0.6)]~{\rm MeV} =
(7.4 \pm 1.2)~{\rm MeV}~.
\eeq
The violation of this sum rule is further evidence that one cannot
always assume equal values of $\Delta,a,c$ for bottom and lighter-quark
systems.

\section{CHARM -- BEAUTY RELATIONS \label{sec:bc}}

\subsection{Universal parameters?}

The comparison of isospin-violating parameters among light-quark, charmed, and
beauty hadrons shows that one cannot regard them as universal.  Suppose,
first of all, that one took $\Delta^m = \Delta^m_c = \Delta^m_b$.  With this
assumption one could solve Eqs.~(\ref{eqn:Dacm}) and (\ref{eqn:bm}) to obtain
$a^m_c = -0.54$ MeV, $a^m_b = 11.42$ MeV.  This makes little sense because
the parameter $a^m_c$ should be positive.

One could, instead, assume that the heavy-quark parameters
\beq
\Delta^m_Q \equiv \Delta^m_c = \Delta^m_b~,~~a^m_Q \equiv a^m_c = a^m_b
\eeq
are equal for charmed and beauty mesons.  (As we shall see, this is
approximately true for baryons.) Then solving Eqs.\ (\ref{eqn:Dacm})
and (\ref{eqn:Dabm}), assuming $h^m_b = 0$, one finds
\beq
\Delta^m_Q = -1.46~{\rm MeV}~,~~a^m_Q = 3.45~{\rm MeV}~,
\eeq
to be compared with the light-quark meson value (see Sec.\ III A)
\beq
\Delta^m = -4.117~{\rm MeV}~,~~a^m = 2.119~{\rm MeV}~.
\eeq
The larger value of $a$ makes sense, because of deeper binding of charmed and
bottom hadrons (hence a larger expectation value of $1/r$).  However, the
difference between $\Delta^m_Q$ [close to the value in Eq.\
(\ref{eqn:sbfit})] and $\Delta^m$ is somewhat puzzling.  Note
that in Table IV we found $\Delta = -2.49$ MeV for light-quark baryons,
considerably different from the value $\Delta^m$.

\subsection{Relations between hyperfine splittings}

Although it is not an isospin splitting, a relation between charmed meson and
beauty meson hyperfine splittings makes used of the relatively new result
from the BaBar Collaboration \cite{TheBaBar:2017yff} which enters the Particle
Data Group compilation.  The relation \cite{Rosner:1992qw} (updated in Ref.\
\cite{Goity:2007fu} to account for QCD corrections) is
\beq \label{eqn:cbhf}
M(\bar B^*_s) - M(\bar B_s) - [M(\bar B^{*0}) - M(\bar B^0)] = (m_c/m_b)
\left\{M(D^*_s) - M(D_s) - [M(D^{*+} - M(D^+)] \right\}
\eeq
The left- and right-hand sides of this equation, based on heavy-quark symmetry,
are related to one another by $b \leftrightarrow c$.  The present status of its
terms is summarized in Table \ref{tab:cbhf} (\cite{Tanabashi:2018oca}).
\begin{table}
\caption{Masses (in MeV) contributing to relation (\ref{eqn:cbhf}) between
charmed and beauty meson hyperfine splittings.
\label{tab:cbhf}}
\begin{center}
\begin{tabular}{c c} \hline \hline
State & Mass \\ \hline
\upstrut
 $\bar B^*_s$  & $5415.4^{+1.8}_{-1.5}$ \\
  $\bar B_s$   & $5366.88 \pm 0.17$ \\
$\bar B^*_s - \bar B_s$ & $48.6^{+1.8}_{-1.5}$ \\
 $\bar B^{*0}$ & $5324.70 \pm 0.22^a$ \\
  $\bar B^0$   & $5279.63 \pm 0.15$ \\
$\bar B^{*0} - \bar B^0$ & $45.07 \pm 0.21^b$ \\
   $D^*_s$     &  $2112.2 \pm 0.4$  \\
    $D_s$      & $1968.34 \pm 0.07$ \\
 $D^*_s - D_s$ & $143.86 \pm 0.41$ \\
   $D^{*+}$    & $2010.26 \pm 0.05$ \\
    $D^+$      & $1869.65 \pm 0.05$ \\
$D^{*+} - D^+$ & $140.603 \pm 0.015$ \\ \hline \hline
\end{tabular}
\end{center}
\leftline{$^a$The charge of the state is not specified in Ref.\
\cite{Tanabashi:2018oca}.}
\leftline{\strut\kern1.3ex
Instead, we quote value for production-weighted average.}
\leftline{$^b$Estimate based on small isospin splitting between charged and
neutral $\bar B^*$.}
\end{table}
The left-hand side of Eq.\ (\ref{eqn:cbhf}) is $3.5 \pm 1.7$ MeV, while the
right-hand side is $(m_c/m_b)(3.26 \pm 0.41)$ MeV $\simeq (1.09 \pm 0.14)$
MeV.  A decisive test of this relation awaits separate measurements of the
masses of $\bar B^{*+}$ and $\bar B^{*0}$, and a reduced error on the mass
of $B^*s$.

\section{COMPARISON WITH OTHER APPROACHES \label{sec:comp}}

Thanks to improvements in computing power, lattice quantum chromodynamics (LQCD)
is beginning to be able to take into account isospin splittings in masses and
decay constants.  (For some references on the latter, see \cite{fm}.)  For
LQCD approaches to light-quark splittings see Refs.\ \cite{Borsanyi:2013lga,%
Borsanyi:2014jba} (octet baryons) and \cite{Horsley:2015eaa} (octet mesons and
baryons).  We look forward to LQCD calculations of isospin splittings in
mesons and baryons containing at least one heavy quark.  

Within quark models there is a long history of tackling isospin splittings in
hadrons \cite{Rosner:1998zc,Brodsky:2011zs,Karliner:2017gml,Isgur:1979ed,%
Itoh:1978ak,Franklin:1981et,Chan:1985ty,Tiwari:1985ru,Hwang:1986ee,%
Capstick:1987cw,Verma:1988gg,Cutkosky:1993cc,Franklin:1995hc,Varga:1998wp,%
SilvestreBrac:2003kd,Durand:2005tb,Ha:2007zz,Ha:2007av,Hwang:2008dj,%
Fritzsch:2008mf,Guo:2008ns,Dillon:2009pf}. (The second-to-last reference,
though using chiral perturbation theory, gives an extensive list of works based
on quark models.)  The parameters $\Delta$ (or $m_u - m_d$) and $a$, when
given, are compared in Table \ref{tab:comp}.  We show there also the latest
estimate of $m_u - m_d$ in the current-quark picture \cite{Tanabashi:2018oca}.

\begin{table}
\caption{Comparison of parameters governing isospin splittings in
quark models.
\label{tab:comp}}
\begin{center}
\begin{tabular}{c c c c} \hline \hline
Reference & $\Delta$ or $m_u-m_d$ &  $a$  &  Comments \\
          &            (MeV)      & (MeV) &           \\ \hline
\upstrut
This work & $\Delta^m = -4.117$ & $a^m = 2.119$ & Light-quark meson octet \\
          & $\Delta^b = -2.491$ & $a^b = 3.052$ & Light-quark baryons
\downstrut \\
\hline
\upstrut
\cite{Rosner:1998zc} & $\Delta^b = -2.57^a$ & $a^b = 3.06^a$ & Neglecting
  kinetic term $K$ \\
\hline
\upstrut
\cite{Brodsky:2011zs}   &                      & $a^m = 3.18 \pm 0.48$ &
 Eq.\ (15) and Appendix A \\
\hline
\upstrut
\cite{Karliner:2017gml} & $\Delta^b = -2.48^a$ & $a^b = 3.05^a$ & \\
                        & $\Delta^b = -2.67^b$ & $a^b = 2.83^b$ & \\
\hline
\upstrut
\cite{Tanabashi:2018oca} & $m_u - m_d = {-}2.55\pm 0.25$ && 
$\overline{\vrule width 0pt height 1.8ex \hbox{MS}}$,
$\mu_{\hbox{\scriptsize renorm.}} =2$ GeV
\\ \hline
\upstrut
\cite{Isgur:1979ed} & $m_u-m_d = -6$ & & \\
\hline
\upstrut
\cite{Itoh:1978ak} & $m_u-m_d = -3.8$ & & \\
\hline
\upstrut
\cite{Franklin:1981et} & $m_u-m_d = -2.54 \pm 0.04$ & & J. Franklin, priv.\
commun.\ \\
\hline
\upstrut
\cite{Chan:1985ty} & $m_u-m_d = -2.66$ & $a^m = 1.5 \pm 0.5$ & Baryon $a$
 unclear \\
\hline
\upstrut
\cite{Tiwari:1985ru} & $m_u-m_d = -4.12$ & & MIT bag model \\ 
\hline
\upstrut
\cite{Hwang:1986ee} & $m_u-m_d = -6.7$ & & MIT bag model \\
\hline
\upstrut
\cite{Capstick:1987cw} & $m_u-m_d = -4.4$ & $a^b = 2.9$ & \\
\hline
\upstrut
\cite{Verma:1988gg} & $m_u-m_d = -2.4$ & Ignored & ``Photon cloud'' effects \\
\hline
\upstrut
\cite{Varga:1998wp} & $m_u-m_d = -11$ & 3.39 & Potential models \\
\hline
\upstrut
\cite{Durand:2005tb} & $m_u-m_d = -1.88$ & 3.52 & Including 3-body terms \\
\hline
\upstrut
\cite{Ha:2007zz} & $m_u-m_d = -1.82$ & & \\
\hline
\upstrut
\cite{Hwang:2008dj} & $\Delta^b = -1.84\pm0.16$ & & \\
\hline
\upstrut
\cite{Fritzsch:2008mf} & $m_u-m_d = -2.5$ & & \\
\hline
\upstrut
\cite{Gasser:1982ap} & $m_u - m_d = {-} 5.7$ &&
$\overline{\vrule width 0pt height 1.8ex \hbox{MS}}$,
$\mu_{\hbox{\scriptsize renorm.}} = 100$ MeV \\
                     & $m_u - m_d = {-} 4.7$ &&
$\overline{\vrule width 0pt height 1.8ex \hbox{MS}}$,
$\mu_{\hbox{\scriptsize renorm.}} = 200$ MeV \\
\hline \hline
\end{tabular}
\end{center}
\leftline{$^a$Different masses for quarks in mesons and baryons}
\leftline{$^b$Universal masses for quarks in mesons and baryons}
\end{table}

The relation between current-quark masses (see the mini-review No.\ 66
in Ref.\ \cite{Tanabashi:2018oca}, and the formalism set forth in Ref.\
\cite{Gasser:1982ap}) and the constituent-quark masses we are using has
been discussed in \cite{Isgur:1979ed}.  However, it has been pointed out
in \cite{Tanabashi:2018oca} that this relation (and hence the definition
of constituent-quark masses) is model-dependent.  We note that many of our
determinations of $m_u-m_d$ in the constituent-quark picture are not that
far from the current-quark value of $\sim -2.5$ MeV,\footnote{At a scale of 2
GeV, one recent lattice QCD determination \cite{Bazavov:2018omf} finds $m_u =
2.130(41)$ MeV, $m_d = 4.675(56)$ MeV, while another \cite{Giusti:2017dmp}
finds $m_u = 2.50 \pm 0.17$ MeV, $m_d = 4.88 \pm 0.20$ MeV.} suggesting that in
those cases the QCD ``dressing'' of current quarks may act linearly on their
masses.  (An exception is presented by the light-quark mesons, for which
$|m_u-m_d|$ is considerably larger, and by the strange-quark mass, which is
about 90 MeV heavier than the average non-strange mass in the current-quark
picture \cite{Tanabashi:2018oca,Bazavov:2018omf} but 180 MeV heavier than the
average non-strange mass in our constituent-quark picture (see the caption of
Table \ref{tab:fit}).

\strut\vskip-1.4cm\strut
\section{CONCLUSIONS \label{sec:concl}}

Within a constituent-quark picture, we have updated predictions of isospin
splittings in hadrons with at most one $c$ or $b$ quark.  Effects considered
included an intrinsic $u$--$d$ mass difference and its effect on kinetic
energies (parameter $\Delta$), Coulomb interactions among the constituent
quarks (parameter $a$), and quark mass dependence on strong and electromagnetic
hyperfine splittings (parameters $b$ and $c$, respectively).  The parameter
$\Delta$ is found to have a non-universal value, ranging from $-4.1$ MeV in
light-quark mesons to $-1.5$ MeV in heavy-quark mesons and possibly in
$b$-quark baryons.  This latter conclusion is preliminary in the
absence of a direct measurement of the masses of both $B^*$ charge states.
A value of $\Delta$ near --2.5 MeV seems consistent with isospin splittings
in light-quark and charmed baryons, but more negative than in bottom
baryons.  Most estimates of the Coulomb interaction term $a$ lie between
2 and 3 MeV.

Quantities whose measurement would help to test relations in the present
analysis include improved masses of $K^{*\pm}$ and $B_s^*$; some isospin
splittings in beauty baryons; and $M(\Xi_{cc}^{++}) - M(\Xi_{cc}^+)$, predicted
in Ref.\ \cite{Karliner:2017gml} to be $(2.17 \pm 0.11)$ MeV under the present
set of assumptions [and $(1.49 \pm 0.12)$ MeV in a model with universal quark
masses for mesons and baryons.] We look forward to these developments,
summarized in Table \ref{tab:summ}, in the data.
\bigskip

\begin{table}
\caption{Observables needed to refine understanding of isospin breaking.
\label{tab:summ}}
\begin{center}
\begin{tabular}{c c} \hline \hline
\upstrut
Observable & Value, if known
\\ \hline\hline
\upstrut
$M(B^{*0})$ & {\bf --}
\\ \hline
$M(K^{*\pm})$ & $891.76\pm0.25$ MeV, hadroproduction
\\
& $895.5\phantom{0}\pm0.8\phantom{5}$ MeV, $\tau$ decay\kern9ex\strut
\\ \hline
\upstrut
$M(B_s^*)$ & $5415.4^{+1.8}_{-1.5}$ MeV \\
\hline
$\Sigma_2^* \equiv
M(\Sigma^{*+}) - 2 M(\Sigma^{*0}) + M(\Sigma^{*+})$ &
  $\kern0.45em 2.6 \pm 2.1$\kern0.5em MeV \\ \hline
$M(\Xi_{cc}^{++}) - M(\Xi_{cc}^+)$ & Predicted in Ref.\ \cite{Karliner:2017gml}
\\ \hline\hline
\end{tabular}
\end{center}
\end{table}

\section*{ACKNOWLEDGMENTS}

We thank Tom DeGrand for a question leading to the present investigation;
Mike Sokoloff for alerting us to Refs.\ \cite{TheBaBar:2017yff} and
\cite{Goity:2007fu} and for a useful conversation; and Davide Giusti and
Feng-Kun Guo for helpful comments on the manuscript.


\begin{thebibliography}{99}

\bibitem{Rosner:1992qw} J.~L.~Rosner and M.~B.~Wise,
  ``Meson masses from SU(3) and heavy quark symmetry,''
  Phys.\ Rev.\ D {\bf 47}, 343 (1993).

\bibitem{Rosner:1998zc} J.~L.~Rosner,
  ``Improved tests of relations for baryon isomultiplet splittings,''
  Phys.\ Rev.\ D {\bf 57}, 4310 (1998) [hep-ph/9707473].

\bibitem{Brodsky:2011zs}
  S.~J.~Brodsky, F.~K.~Guo, C.~Hanhart and U.~G.~Meissner,
  ``Isospin splittings of doubly heavy baryons,''
  Phys.\ Lett.\ B {\bf 698}, 251 (2011) [arXiv:1101.1983 [hep-ph]].

\bibitem{Karliner:2017gml} M.~Karliner and J.~L.~Rosner,
  ``Isospin splittings in baryons with two heavy quarks,''
  Phys.\ Rev.\ D {\bf 96}, 033004 (2017) [arXiv:1706.06961 [hep-ph]].

\bibitem{TheBaBar:2017yff} J.~P.~Lees {\it et al.} (BaBar Collaboration),
  ``Measurement of the $D^{*+}(2010)-D^+$ Mass Difference,''
  Phys.\ Rev.\ Lett.\ {\bf 119}, 202003 (2017) [arXiv:1707.09328 [hep-ex]].

\bibitem{Tanabashi:2018oca} M.~Tanabashi {\it et al.} (Particle Data Group),
  ``Review of Particle Physics,''
  Phys.\ Rev.\ D {\bf 98}, 030001 (2018).

\bibitem{Goity:2007fu} J.~L.~Goity and C.~P.~Jayalath,
  ``Strong and Electromagnetic Mass Splittings in Heavy Mesons,''
  Phys.\ Lett.\ B {\bf 650}, 22 (2007) [hep-ph/0701245].

\bibitem{Fanti:1999gy} V.~Fanti {\it et al.} (NA48 Collaboration), ``Precision
  measurement of the $\Xi^0$ mass and the branching ratios of the decays
  $\Xi^0 \to \Lambda \gamma$ and $\Xi^0 \to \Sigma^0 \gamma$,''
  Eur.\ Phys.\ J.\ C {\bf 12}, 69 (2000).

\bibitem{DeRujula:1975qlm} A.~De R\'ujula, H.~Georgi and S.~L.~Glashow,
  ``Hadron Masses in a Gauge Theory,'' Phys.\ Rev.\ D {\bf 12}, 147 (1975).

\bibitem{Lipkin:1978eh} H.~J.~Lipkin,
  ``A Unified Description of Mesons, Baryons and Baryonium,''
  Phys.\ Lett.\ {\bf 74B}, 399 (1978).

\bibitem{Gasiorowicz:1981jz} S.~Gasiorowicz and J.~L.~Rosner,
  ``Hadron Spectra and Quarks,''
  Am.\ J.\ Phys.\ {\bf 49}, 954 (1981).

\bibitem{Karliner:2014gca} M.~Karliner and J.~L.~Rosner,
  ``Baryons with two heavy quarks: Masses, production, decays, and detection,''
  Phys.\ Rev.\ D {\bf 90}, 094007 (2014) [arXiv:1408.5877 [hep-ph]].

\bibitem{Aaij:2017ueg} R.~Aaij {\it et al.} (LHCb Collaboration),
  ``Observation of the doubly charmed baryon $\Xi_{cc}^{++}$,''
  Phys.\ Rev.\ Lett.\ {\bf 119}, 112001 (2017) [arXiv:1707.01621 [hep-ex]].

\bibitem{Karliner:2016zzc} M.~Karliner, S.~Nussinov and J.~L.~Rosner,
  ``$Q Q \bar Q \bar Q$ states: masses, production, and decays,''
  Phys.\ Rev.\ D {\bf 95}, 034011 (2017) [arXiv:1611.00348 [hep-ph]].

\bibitem{Isgur:1979ed} N.~Isgur, ``Isospin violating mass differences and
  mixing angles: the role of quark masses,'' Phys.\ Rev.\ D {\bf 21}, 779
  (1980) Erratum: [Phys.\ Rev.\ D {\bf 23}, 817 (1981)].

\bibitem{CG} S. Coleman and S. L. Glashow, ``Departures from the eightfold
   way: Theory of strong interaction symmetry breakdown,''
   Phys.\ Rev.\ {\bf 134}, B671 (1964). 

\bibitem{Franklin:1975yu} J.~Franklin,
  ``Mass Relations for Charmed Baryons,''
  Phys.\ Rev.\ D {\bf 12}, 2077 (1975).

\bibitem{Aaij:2014esa} R.~Aaij {\it et al.} (LHCb Collaboration),
  ``Precision measurement of the mass and lifetime of the $\Xi_b^0$ baryon,''
  Phys.\ Rev.\ Lett.\ {\bf 113}, 032001 (2014) [arXiv:1405.7223 [hep-ex]].

\bibitem{Aaij:2014lxa} R.~Aaij {\it et al.} (LHCb Collaboration),
  ``Precision Measurement of the Mass and Lifetime of the $\Xi_b^-$ Baryon,''
  Phys.\ Rev.\ Lett.\  {\bf 113}, 242002 (2014) [arXiv:1409.8568 [hep-ex]].

\bibitem{fm} J. L. Rosner, S. L. Stone, and R. S. Van de Water, mini-review
of leptonic decays of charged pseudoscalar mesons, in \cite{Tanabashi:2018oca},
p.\ 700.

\bibitem{Borsanyi:2013lga} 
  S.~Borsanyi {\it et al.} [Budapest-Marseille-Wuppertal Collaboration],
  ``Isospin splittings in the light baryon octet from lattice QCD and QED,''
  Phys.\ Rev.\ Lett.\ {\bf 111}, 252001 (2013) [arXiv:1306.2287 [hep-lat]].

\bibitem{Borsanyi:2014jba} S.~Borsanyi {\it et al.},
  ``Ab initio calculation of the neutron-proton mass difference,''
  Science {\bf 347}, 1452 (2015) [arXiv:1406.4088 [hep-lat]].

\bibitem{Horsley:2015eaa} R.~Horsley {\it et al.},
  ``Isospin splittings of meson and baryon masses from three-flavor lattice QCD + QED,''
  J.\ Phys.\ G {\bf 43}, 10LT02 (2016) [arXiv:1508.06401 [hep-lat]].

\bibitem{Itoh:1978ak} C.~Itoh, T.~Minamikawa, K.~Miura and T.~Watanabe,
  ``Electromagnetic Mass Differences of Lowest Lying Hadrons and Behavior of Quarks in Hadrons,''
  Prog.\ Theor.\ Phys.\ {\bf 61}, 548 (1979).

\bibitem{Franklin:1981et} J.~Franklin and D.~B.~Lichtenberg,
  ``Estimate of the Quark - Gluon Coupling Strength From Baryon Masses,''
  Phys.\ Rev.\ D {\bf 25}, 1997 (1982).

\bibitem{Chan:1985ty} L.~H.~Chan,
  ``Isospin Mass Splittings of Hadrons With Heavy Quarks,''
  Phys.\ Rev.\ D {\bf 31}, 204 (1985).

\bibitem{Tiwari:1985ru} K.~P.~Tiwari, C.~P.~Singh and M.~P.~Khanna,
  ``Electromagnetic Mass Splittings Of Heavier Hadrons In The Mit Bag Model,''
  Phys.\ Rev.\ D {\bf 31}, 642 (1985).

\bibitem{Hwang:1986ee} W.~Y.~P.~Hwang and D.~B.~Lichtenberg,
  ``Mass Splitting of Heavy Baryon Isospin Multiplets,''
  Phys.\ Rev.\ D {\bf 35}, 3526 (1987).

\bibitem{Capstick:1987cw} S.~Capstick, ``Isospin Violations in Baryons and the
  $\Sigma_c^0 - \Sigma_c^{++}$ Mass Difference,''
  Phys.\ Rev.\ D {\bf 36}, 2800 (1987).

\bibitem{Verma:1988gg} R.~C.~Verma and S.~Srivastava, ``Photon Cloud Effects on
  Isomultiplet Mass Differences of Charmed and Uncharmed Baryons,''
  Phys.\ Rev.\ D {\bf 38}, 1623 (1988).

\bibitem{Cutkosky:1993cc} R.~E.~Cutkosky and P.~Geiger,
  ``Isospin splitting in heavy baryons and mesons,''
  Phys.\ Rev.\ D {\bf 48}, 1315 (1993) [hep-ph/9304202].

\bibitem{Franklin:1995hc} J.~Franklin, ``Sum rules for charmed baryon masses,''
  Phys.\ Rev.\ D {\bf 53}, 564 (1996) [hep-ph/9506318].

\bibitem{Varga:1998wp} K.~Varga, M.~Genovese, J.~M.~Richard, and
 B.~Silvestre-Brac,
  ``Isospin mass splittings of baryons in potential models,''
  Phys.\ Rev.\ D {\bf 59}, 014012 (1999) [hep-ph/9803340].

\bibitem{SilvestreBrac:2003kd} B.~Silvestre-Brac, F.~Brau and C.~Semay,
  ``Electromagnetic splitting for mesons and baryons using dressed constituent
  quarks,'' J.\ Phys.\ G {\bf 29}, 2685 (2003) [hep-ph/0302252].

\bibitem{Durand:2005tb} L.~Durand and P.~Ha,
  ``Electromagnetic corrections to baryon masses,''
  Phys.\ Rev.\ D {\bf 71}, 073015 (2005)
  Erratum: [Phys.\ Rev.\ D {\bf 75}, 039903 (2007)] [hep-ph/0502090].

\bibitem{Ha:2007zz} P.~Ha,
  ``Estimates of isospin breaking contributions to baryon masses,''
  Phys.\ Rev.\ D {\bf 76}, 073004 (2007).

\bibitem{Ha:2007av} P.~Ha,
  ``A Parametrization of the baryon octet and decuplet masses,''
  J.\ Phys.\ G {\bf 35}, 075006 (2008) [arXiv:0711.4364 [hep-ph]].

\bibitem{Hwang:2008dj} C.~W.~Hwang and C.~H.~Chung,
  ``Isospin mass splittings of heavy baryons in HQS,''
  Phys.\ Rev.\ D {\bf 78}, 073013 (2008) [arXiv:0804.4044 [hep-ph]].

\bibitem{Fritzsch:2008mf} H.~Fritzsch,
  ``Isospin Mass Differences of Heavy Baryons,'' arXiv:0811.0481 [hep-ph].

\bibitem{Guo:2008ns} F.~K.~Guo, C.~Hanhart and U.~G.~Meissner, ``Mass
  splittings within heavy baryon isospin multiplets in chiral perturbation
  theory,'' JHEP {\bf 0809}, 136 (2008) [arXiv:0809.2359 [hep-ph]].

\bibitem{Dillon:2009pf} G.~Dillon and G.~Morpurgo,
  ``The General QCD parametrization and the hierarchy of its parameters: Why
  some simple models of hadrons work so well,''
  Riv.\ Nuovo Cim.\ {\bf 33}, 1 (2010) [arXiv:0910.5326 [hep-ph]].

\bibitem{Gasser:1982ap} J.~Gasser and H.~Leutwyler, ``Quark Masses,''
  Phys.\ Rept.\ {\bf 87}, 77 (1982).

\bibitem{Bazavov:2018omf} A.~Bazavov {\it et al.} (Fermilab Lattice and MILC
  and TUMQCD Collaborations), ``Up-, down-, strange-, charm-, and bottom-quark
  masses from four-flavor lattice QCD,''
  Phys.\ Rev.\ D {\bf 98}, 054517 (2018) [arXiv:1802.04248 [hep-lat]].

\bibitem{Giusti:2017dmp} D.~Giusti, V.~Lubicz, C.~Tarantino, G.~Martinelli,
  S.~Sanfilippo, S.~Simula and N.~Tantalo, ``Leading isospin-breaking
  corrections to pion, kaon and charmed-meson masses with Twisted-Mass
  fermions,'' Phys.\ Rev.\ D {\bf 95}, 114504 (2017)
  [arXiv:1704.06561 [hep-lat]].

\end{thebibliography}
\end{document}